\documentclass[prd,aps,superscriptaddress,amsmath,amssymb, nobibnotes,nofootinbib,twocolumn,10pt]{revtex4-1}
\usepackage{graphicx} 

\usepackage[colorlinks=true, citecolor=red,linkcolor=blue,urlcolor=magenta]{hyperref}
\usepackage{tikz}

\usepackage{soul}
\DeclareMathOperator{\Tr}{Tr}
\newcommand{\nc}{\newcommand}
\nc{\n}{\textrm{n}}
\usepackage{xcolor}

\begin{document}

\title{Entanglement negativity between separated regions in quantum critical systems}
\author{Gilles Parez}
\affiliation{\it D\'epartement de Physique, Universit\'e de Montr\'eal, Montr\'eal, QC H3C 3J7, Canada}
\affiliation{\it Centre de Recherches Math\'ematiques, Universit\'e de Montr\'eal, Montr\'eal, QC H3C 3J7, Canada}

\author{William Witczak-Krempa}
\affiliation{\it D\'epartement de Physique, Universit\'e de Montr\'eal, Montr\'eal, QC H3C
3J7, Canada}
\affiliation{\it Centre de Recherches Math\'ematiques, Universit\'e de Montr\'eal, Montr\'eal, QC H3C 3J7, Canada}

\affiliation{\it Institut Courtois, Universit\'e de Montr\'eal, Montr\'eal, QC H2V 0B3, Canada}

\begin{abstract}
We study the entanglement between disjoint subregions in quantum critical systems through the lens of the logarithmic negativity. We work with systems in arbitrary dimensions, including conformal field theories and their corresponding lattice Hamiltonians, as well as resonating valence-bond states. At small separations, the logarithmic negativity is big and displays universal behavior, but we show non-perturbatively that it decays faster than any power at large separations. This can already be seen in the minimal setting of single-spin subregions. 
The corresponding absence of distillable entanglement at large separations generalizes the 1d result, and indicates that quantum critical groundstates do not possess long-range bipartite entanglement, at least for bosons. For systems with fermions, a more suitable definition of the logarithmic negativity exists that takes into account fermion parity, and we show that it decays algebraically. Along the way we obtain general 
results for the moments of the partially transposed density matrix.
\end{abstract}

\maketitle

\section{Introduction}
Quantum critical groundstates, such as the $d$-dimensional transverse-field Ising model at its transition, possess algebraically decaying correlation functions at large separations. 
It is thus natural to ask whether this translates to the amount of entanglement between separated subregions. Remarkably, for general bosonic conformal field theories (CFTs) in one spatial dimension, it was shown that this is not the case: the logarithmic negativity (LN) \cite{VW02}, which is a proper entanglement monotone \cite{plenio2005logarithmic} contrary to the entanglement entropy, decays faster than any power \cite{CCT12, CCT13}. In higher dimensions, this was shown to hold for the non-interacting scalar field~\cite{klco2021entanglement,klco2021geometric} as well as for resonating valence-bond (RVB) states~\cite{parez2023separability} on arbitrary lattices (in agreement with continuum results~\cite{Boudreault:2021pgj}). But can it be true for all highly correlated 
quantum critical systems in any dimension?  

We show that the answer is yes by using (i) exact properties of the 2-spin reduced density matrix in the $d$-dimensionsal quantum Ising model, and (ii) a non-perturbative expansion of twist operators in general CFTs. However, it is not the end of the story for  
systems that contain local fermion operators, such as massless Majorana or Dirac fermions. For such fermionic theories another definition of the LN was introduced \cite{SSR17} that takes into account the fact that physical density matrices must respect fermion parity. In fermionic
quantum critical systems, we find that this alternative LN, a proper entanglement monotone for fermions \cite{shapourian2019entanglement}, decays algebraically.

\section{Entanglement negativity}
\begin{figure}
    \centering
    
    \begin{tikzpicture}[>=stealth, scale = 0.5]
  
  \draw [ ultra thick] (-0.6,10) node[scale=0.7] {\large \textbf{a)}};
      \draw [ ultra thick] (5.4,10) node[scale=0.7] {\large \textbf{b)}};
  	
  		\draw [thick,stealth-stealth] (2,3.9) -- (2,6.1);
  		\draw [ ultra thick] (1.6,5) node[scale=0.6] {\huge ${r}$};
  		
  		\draw [thick, stealth-stealth] (-0.1,0.2) -- (-0.1,3.8);
  		\draw [ ultra thick] (-0.5,2) node[scale=0.6] {\huge ${\ell}$};
  
  	 \draw [ blue, fill=blue!25, opacity=0.5] (0.2, 6.2) rectangle (3.8,9.8);
  	 \draw [ ultra thick] (2,8) node[scale=0.6] {\huge ${ A_1}$};
  	 
  	 \draw [ red, fill=red!25, opacity=0.5] (0.2, 0.2) rectangle (3.8,3.8);
  	 \draw [ ultra thick] (2,2) node[scale=0.6] {\huge ${ A_2}$};
  	 
   \draw [] (6,9) -- (8,9);
  	 \draw [] (8.7,9) -- (10.7,9);  	 
  	 
  	 \draw [] (6,8) -- (8,8);
  	 \draw [] (8.7,8) -- (10.7,8);
  	 
  	  \draw [] (6,7) -- (8,7);
  	 \draw [] (8.7,7) -- (10.7,7);
  	 
  	 \draw [-stealth] (6,6.5) -- (7.3,6.5);
  	 \draw [  thick] (6.65,6.2) node[scale=.9] { ${\tau}$};
  	 
  	\draw [  thick] (8.1,6.5) node[scale=.9] { ${0^-}$};
  	\draw [  thick] (8.9,6.5) node[scale=.9] { ${0^+}$};

  	 \draw [orange] (8,9) -- (8.7,8);
  	 \draw [orange] (8,8) -- (8.7,7);
  	 \draw [orange, dashed] (8,7) -- (8.7,9);
  	 
  	  \draw [ blue, fill=blue!25, opacity=1] (8, 9) circle (2pt);
  	 \draw [ blue, fill=blue!25, opacity=1] (8.7, 9) circle (2pt);
  	 \draw [ blue, fill=blue!25, opacity=1] (8, 8) circle (2pt);
  	 \draw [ blue, fill=blue!25, opacity=1] (8.7, 8) circle (2pt);
  	 \draw [ blue, fill=blue!25, opacity=1] (8, 7) circle (2pt);
  	 \draw [ blue, fill=blue!25, opacity=1] (8.7, 7) circle (2pt);

  	 	  \draw [] (6,3) -- (8,3);
  	 \draw [] (8.7,3) -- (10.7,3);  	 
  	 
  	 \draw [] (6,2) -- (8,2);
  	 \draw [] (8.7,2) -- (10.7,2);
  	 
  	  \draw [] (6,1) -- (8,1);
  	 \draw [] (8.7,1) -- (10.7,1);

  	 \draw [orange] (8,1) -- (8.7,2);
  	 \draw [orange] (8,2) -- (8.7,3);
  	 \draw [orange, dashed] (8,3) -- (8.7,1);  	   	   	 
  	 \draw [ red, fill=red!25, opacity=1] (8, 1) circle (2pt);
  	  \draw [ red, fill=red!25, opacity=1] (8.7, 1) circle (2pt);
  	  
  	   \draw [ red, fill=red!25, opacity=1] (8, 2) circle (2pt);
  	  \draw [ red, fill=red!25, opacity=1] (8.7, 2) circle (2pt);
  	  
  	   \draw [ red, fill=red!25, opacity=1] (8, 3) circle (2pt);
  	  \draw [ red, fill=red!25, opacity=1] (8.7, 3) circle (2pt);
  
  \end{tikzpicture} 
 \caption{a) Subregion $A$ is the union of two disjoint parts $A_1$ and $A_2$, here two squares in $d=2$. 
 b) Replicated space-time used to evaluate the moments of the partially transposed density matrix $\mathcal E_n^b$ for $n=3$. We show the flow of time for a trajectory that traverses the three copies of $A_1$ (top), and $A_2$ (bottom). The partial transposition on $A_1$ time-reverses the trajectory relative to $A_2$.}
    \label{fig:geo}
\end{figure}
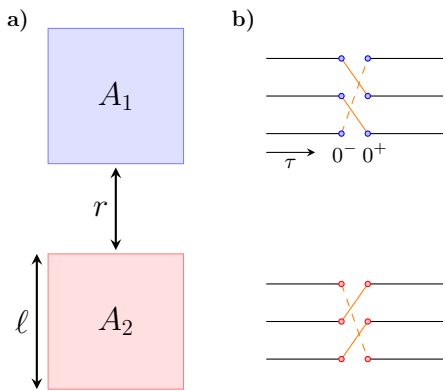

Consider the groundstate of a quantum critical system (such as a CFT), and $A=A_1A_2$ is a subregion made of two disjoint parts, see Fig.~\ref{fig:geo}. Tracing out the complement of $A$ gives the reduced density matrix $\rho_A$. The bosonic ($b$) \cite{Zyczkowski:1998yd,Eisert:1998pz,VW02} and fermionic ($f$) \cite{SSR17} logarithmic negativities read
\begin{align} \label{eq:ln}
    \mathcal E^{b/f}\mkern-2.4mu (A_1,A_2) = \log \big\lVert \rho_A^{T_1^{b/f}} \!\big\rVert_1
\end{align}
where $\lVert X\rVert_1=\Tr \sqrt{XX^\dag}$ is the trace norm. 
The bosonic partial transpose (PT) $T_1^b$ swaps the entries of the density matrix pertaining to subregion $A_1$ only,
\begin{align} \label{eq:pt}
    T_1^b\left( |\alpha_1\rangle|\beta_2\rangle\, \langle\widetilde\alpha_1|\langle\widetilde\beta_2|\right)=|\widetilde\alpha_1\rangle|\beta_2\rangle\, \langle\alpha_1|\langle\widetilde\beta_2|,
\end{align}
where the subscript denotes the subregion supporting the state. One can use \eqref{eq:pt} for fermionic states, but this leads to 
many shortcomings. 
For one, if the initial state is Gaussian, the partially transposed one is generally not Gaussian \cite{eisler2015partial}, making the calculation of $\mathcal E^b$ arduous even in the simplest cases. Second, the LN defined via \eqref{eq:pt} fails to detect topological phases \cite{SSR17}. One reason behind these issues is that $T_1^b$ does not account for the anti-commutation relations between fermions. 
One can define an alternative \emph{fermionic} partial transpose $T_1^f$ that removes the above shortcomings \cite{SSR17}. In the number-occupation basis, the fermionic partial transpose acts as~\eqref{eq:pt} but it incorporates an additional phase factor that depends on the occupation numbers, see example below and the Supplemental Material (SM). 

\section{Scaling with separation}
Let us begin our discussion with the simplest case: $A_1$ and $A_2$ are adjacent. For simplicity, we can take the subregions to be hyper-cubes (such as squares in $d=2$, see Fig.~\ref{fig:geo}) of linear size $\ell$. In that case, both LNs obey the boundary law
\begin{align} \label{eq:bdy}
    \mathcal E^{b/f}\mkern-2.5mu (r=0)= c_{1/2} \left(\frac{\ell}{\delta}\right)^{\! d-1}+\cdots ,
\end{align}
where $\ell^{d-1}$ is the size of the shared boundary, $\delta$ is a UV cutoff such as the lattice spacing, and $c_{1/2}$ is a positive coefficient. Equation~\eqref{eq:bdy} follows from the fact that
when $A$ is the entire space, both LNs reduce to the 1/2-R\'enyi entropy~\cite{VW02,SSR17}, which obeys a boundary law $S_{1/2}(A_1)=c_{1/2}|\partial A_1|/\delta^{d-1}+\dotsb$. Being a UV property, the scaling~\eqref{eq:bdy} 
continues to hold as the subregions shrink so that they are no longer complementary, but $|\partial A_1|$ 
is replaced by the shared boundary. 
For $d=1$, one gets a logarithm instead~\cite{CCT13,shapourian2019twisted}. 

Next, we increase the separation. For finite separations, the LNs are scaling functions of $r/\ell$, and we first study what happens when this ratio approaches zero. In the limit $r\to \delta$
we should recover the boundary law~\eqref{eq:bdy}, so that for small but finite distances, we replace the cutoff by the separation $r$ to obtain the universal scaling
\begin{align} \label{eq:short}
    \mathcal E^{b/f}\mkern-2.5mu  (r\ll \ell)=\kappa^{b/f} \left( \frac{\ell}{r} \right)^{\!d-1} +\cdots ,
\end{align}
where $\kappa^{b/f}$ is a positive coefficient determined by the data of the CFT. For instance, for the 2d free relativistic scalar field, we numerically verified the scaling~\eqref{eq:short} by working with a lattice of coupled quantum harmonic oscillators, and obtained $\kappa^b=0.0226\pm0.0004$.
In 1d, the scaling becomes $\kappa^{b/f}\log(\ell/r)$, where $\ell$ is the length of each of the two intervals. In a fermionic CFT, we now argue that $\kappa_f=\kappa_b$. It is simplest to make the argument on the infinite cylinder, where $A_{1,2}$ are infinite strips separated by a strip of width $r$. If $\kappa^b\neq\kappa^f$, as we take the distance $r\to 0$, $\mathcal E_b-\mathcal E_f$ would diverge as $1/r$. This contradicts the fact that for $A_1$ and $A_2$ adjacent, the LNs
are equal. The difference between the LNs comes from matrix elements that hop into the separating strip, and so these contributions decrease as $r\to 0$. For the 1d free Dirac fermion, we can show equality.
Indeed, for 1d CFTs $\mathcal E^b=\tfrac{c}{4}\log(\ell/r)$ at small $r/\ell$~\cite{CCT13}. The free-fermion central charge is $c\!=\!1$. To evaluate $\mathcal E^f$,
we use the moments of $\rho^{T^f_1}$ obtained for even integers~\cite{shapourian2019twisted}, which we analytically continue to get $\mathcal E^f=\tfrac14\log(\ell/r)$, matching $\mathcal E^b$. Taking $r\!\to\!\delta$, we recover the adjacent case~\cite{CCT12}. 

In the opposite limit of large separations, one possibility is power-law decay since the correlation length is infinite, and correlation functions of all local operators decay algebraically. This is what happens for the mutual information $I(A_1,A_2)$ \cite{casini2009entanglement}, which is a measure of both entanglement and classical correlations \cite{wolf2008area}. However, contrary to the naive expectation, for 1d CFTs, $\mathcal E^b$ decays faster than any power \cite{CCT12}. 
It suggests that sufficiently separated subregions share little (but non-zero~\cite{Verch2004,Hollands2018}) distillable entanglement, at least as quantified by $\mathcal E^b$.  
Beyond 1d, the LN was observed to decay exponentially for the free scalar CFT in 2d and 3d \cite{klco2021entanglement,klco2021geometric}.
We wish to answer the questions: How general is this rapid suppression of entanglement?
Does it hold in fermionic systems when one uses appropriate entanglement measures?

\section{Skeletal approach to negativity}
We begin by considering the simplest case where $A_1$ and $A_2$ are two spins (qubits) that belong to the groundstate of a $d$-dimensional Hamiltonian. 
This extreme skeletal regime \cite{JTBBJH18,berthiere2022entanglement} faithfully captures the absence of power-law scaling in the thermodynamic limit. Indeed, for large separations, the regions become point-like relative to their separation, which justifies fusing operators: this is the celebrated operator-product expansion (OPE). Working with the minimal Hilbert space is tantamount to keeping the dominant operators in the OPE, and we shall show that it gives non-perturbative access to their contributions. An advantage of the skeletal approach is that we obtain the entire density matrix, from which one can extract entanglement measures other than the LN, which becomes essential as the number of spins in $A$ increases.

By tracing out the complement of $A$, one obtains the reduced density matrix in the $S^z$-basis \mbox{$\{|\!\downarrow\downarrow\rangle,|\!\downarrow\uparrow\rangle,|\!\uparrow\downarrow\rangle,|\!\uparrow\uparrow\rangle\}$},
\begin{align} \label{eq:rho}
\rho_A \!=\! \begin{bmatrix}
\tfrac14\!-\!m_z\!+\!\varrho_r^z & 0 & 0 & \varrho_r^x \!-\! \varrho_r^y \\
0 & \tfrac14 \!-\! \varrho_r^z & \varrho_{r}^x \!+\! \varrho_r^y & 0 \\[2pt]
0 & \varrho_r^x \!+\! \varrho_r^y & \tfrac14 \!-\! \varrho_r^z & 0 \\
 \varrho_r^x \!-\! \varrho_r^y & 0 & 0 & \tfrac14 \!+\! m_z \!+\! \varrho_r^z
\end{bmatrix}
\end{align}
where $m_z=\langle S_{i_1}^z\rangle$, and $\varrho_r^\alpha=\langle S_{i_1}^\alpha S_{i_2}^\alpha\rangle$ is the spin-spin correlation function at a separation $r=|i_2-i_1|$ (see SM).
For simplicity, we have assumed that (i) sites $i_1$ and $i_2$ are related by symmetry, and (ii) the groundstate is real in the $S^z$-basis. 
These conditions are not essential, but are respected for instance in the transverse-field Ising model on a general $d$-dimensional lattice, such as hypercubic, $H=-\sum_{\langle i,j\rangle} S_i^x S_j^x -h\sum_i S_i^z$. Regarding the first condition, one can work on the infinite lattice or with periodic boundary conditions; for open boundary conditions, the two sites should be related by reflection. The second condition is respected due to the anti-unitary symmetry given by complex conjugation in the $S^z$-basis, $\mathcal T=\prod_i K_i$, which we call time-reversal. One sees that $\rho_A^{T_1^b}$ is given by $\rho_A|_{\varrho_r^y\to-\varrho_r^y}$, coherent with the fact that time-reversal on $A_1$ yields $\mathcal T_1 S_{i_1}^y \mathcal T_1 \!=\!-S_{i_1}^y$, whereas the $x,z$-components remain invariant and the transformation does not act on $i_2$.  We now examine the fate of entanglement at the quantum critical point $h_c$.
For $r\!\gg\! 1$, $\varrho_r^z\to m_z^2$ approaches a finite constant, while $\varrho_r^{x,y}\to 0$ since the 1-point functions $\langle S_{i}^{x,y}\rangle$ vanish at criticality due to the Ising $\mathbb Z_2$ symmetry. Further, we have the strict inequalities $0<m_z<1/2$ since the 1-point function does not vanish because $S^z_i$ is even under the $\mathbb Z_2$ symmetry, and the groundstate is not fully polarized, which only occurs when $h\to\infty$.  We can therefore accurately approximate $\rho_A^{T_1^b}$ at sufficiently large separations as a diagonal matrix with strictly positive eigenvalues, implying that the LN vanishes. 
Thus, there exists a sudden-death distance $r_{\rm sd}$ at which $\mathcal E^b(r\!\geqslant\! r_{\rm sd})=0$. In particular, the decay is not algebraic. Interestingly, this distance is  small: $r_{\rm sd}=3$ for $d=1$ as was shown using \eqref{eq:rho}~\cite{Osterloh_2002,Osborne,JTBBJH18}. We thus have that $A_1$ and $A_2$ rapidly become un-entangled with separation for all $d$ since 
the positivity of the PT is a necessary \cite{P96} and sufficient \cite{werner2001bound} condition for the separability of two qubits. Furthermore, we show that these properties hold across the entire phase diagram, (SM). 

As a second example, we consider nearest-neighbor $SU(2)$-symmetric RVB states of spins on the square lattice \cite{anderson1973resonating}. These states describe critical phases of matter but do not correspond to CFTs in the continuum limit. Nonetheless, we can also argue that there exists a sudden-death distance in this case. In the skeletal regime, the two-spin density matrix is also given by Eq.~\eqref{eq:rho}, with $\varrho_r^{x} =\varrho_r^{y}= \varrho_r^z$ because of the $SU(2)$ symmetry. Moreover, the spin-spin correlation functions decay exponentially fast \cite{liang1988some,tasaki1989order,albuquerque2010critical,tang2011properties}, $|\varrho_r^z|\sim \exp{(-r/\xi)}$ with $\xi=1.35(1)$~\cite{albuquerque2010critical}. The eigenvalues of the partially transposed density matrix become all positive for a finite values of $\varrho_{r_{\rm sd}}^z$, corresponding to a finite sudden-death distance $r_{\rm sd}$. For larger subregions of two or more spins, the density matrices also depend on dimer-dimer correlations, which decay algebraically \cite{albuquerque2010critical,tang2011properties}. However, the argument still applies and the eigenvalues of the partially transposed density matrix are guaranteed to be positive for small but finite correlations, corresponding to a sudden-death distance. In fact, the argument readily generalizes for RVB states defined on arbitrary lattices, including triangular lattices where the system describes a gapped phase. It is known that the negativity in RVB states on arbitrary lattices is at most exponentially decaying with the separation \cite{parez2023separability}. Our present analysis is consistent but stronger than those results, and further shows that distillable long-range entanglement exactly dies at finite distance in these states.

\section{Negativity moments}

In our analysis below, 
we establish the rapid decrease of $\mathcal E^b$ in the thermodynamic limit in all CFTs for all $d$.
The CFT analysis 
employs the replica trick \cite{CC04}, where one needs to evaluate the normalized moments~\cite{CCT12,CCT13} 
\begin{align}
    \mathcal E_n^b(A_1,A_2) = \log \frac{\Tr\big(\rho_A^{T_1^b}\big)^n}{\Tr \rho_A^n},
\end{align}
with $n=3$ being the first integer that yields a non-zero answer. The $n\to 1$ analytic continuation of the \emph{even} sequence yields the LN $\mathcal E^b$. 
For our 2-qubit case, we can exactly evaluate the moments. We find at large separations $\mathcal E_n^b= b_n \varrho_r^x\varrho_r^y+\cdots$, where the ellipsis denotes subleading terms. 
We have the 
asymptotic relation $\varrho_r^x=A_x/r^{2\Delta_\sigma}$, where $\Delta_\sigma$ is the scaling dimension of the leading $\mathbb Z_2$-odd operator $\sigma$, since $S_i^x$ is the microscopic ferromagnetic order parameter. $S_i^y$ is also $\mathbb Z_2$-odd, but it maps to a CFT operator with a higher scaling dimension, consistent with the strong spin anisotropy of the Ising model. More precisely, $\varrho_r^y=A_y/r^{2(\Delta_\sigma +1)}$, so that the corresponding low-energy operator is the temporal component of the descendant of the order parameter, $\partial_0\sigma$. This can be analytically verified with the exact solution obtained by fermionizing in 1d~\cite{Pfeuty}. We thus have that the moments scale as  
\begin{align} \label{eq:En-power}
    \mathcal E_n^b(r\gg \ell) = B_n\left(\frac{\ell^2}{r^2}\right)^{\! \!2\Delta_\sigma+1},
\end{align}
where we have written the expression in a form that 
holds true in the thermodynamic limit for a large family of CFTs including the $d=1,2$ Ising CFTs, with $\ell$ being the linear size of $A_1$ and $A_2$. The power law that applies to all CFTs, irrespective of their spectra, is given in the next section. 
For the $d=1$ Ising CFT, we have $\Delta_\sigma=1/8$, so that \eqref{eq:En-power} scales as $(\ell/r)^{5/2}$, in agreement
with the continuum result~\cite{A13}. Moreover, we verified that the scaling \eqref{eq:En-power} holds for the free scalar and free Dirac fermion in 1d and 2d. 

For the RVB states on the square lattice, the moments scale as $\mathcal E_n^b(r\gg \ell)\sim \exp{(-2r/\xi)}$ for the 2-spin case. For larger subsystems, the moments also depend on the dimer-dimer, or 4-spin correlation functions, which decay algebraically. These correlations thus dominate the spin-spin ones, and we expect the negativity moments to decay algebraically with the separation. In either cases there is no sudden death for the negativity moments. This argument generalizes to arbitrary lattices.

\section{Field-theoretical analysis}

Consider generic $(d\!+\!1)$-dimensional CFTs, where the subsystems $A_1,A_2$ are regions of $\mathbb{R}^d$ with linear size $\ell$ and separated by a distance~$r$, see Fig.~\ref{fig:geo}.
The trace $\Tr\rho_{A_1 A_2}^n$ corresponds to a normalized partition function in theory where $n$ copies of space are sewed along the cross-section corresponding to imaginary time $\tau=0$ and position ${\bf x} \in A_1 A_2$, as illustrated in the bottom-right panel of Fig.~\ref{fig:geo}. For the trace $\Tr\big(\rho_{A_1 A_2}^{T_1^b}\big)^n$, the sewing for ${\bf x}\in A_1$ connects replicas in the opposite order because of the partial transposition, see the top-right panel of  Fig.~\ref{fig:geo}. The resulting $n$-sheeted spacetimes are denoted $\mathcal{C}_{A_1 A_2}^{(n)}$ and $\mathcal{C}_{A_1 A_2}^{(n)T_1}$, respectively. The sewing can be implemented by non-local defect/twist operators $\Sigma_{A_1 A_2}^{(n)}$ and $\Sigma_{A_1 A_2}^{(n)T_1}$ having support on $A$. For $r\gg \ell$, the twist operators can be factorized into a product of a twist operator acting on $A_1$ and another on $A_2$: $\Tr\rho_{A_1 A_2}^n = \big\langle \Sigma_{A_1}^{(n)}\Sigma_{A_2}^{(n)}\big\rangle_{(\mathbb{R}^{d+1})^n}$, $\Tr\big(\rho_{A_1 A_2}^{T_1^b}\big)^n = \big\langle \Sigma_{A_1}^{(n)T_1}\Sigma_{A_2}^{(n)}\big\rangle_{(\mathbb{R}^{d+1})^n}$. As in Ref.~\cite{C13}, the large-$r$ limit allows us to further expand each twist operator in terms of local operators in the direct product of the CFT, which is the OPE fusion of extended operators,
\begin{equation}
    \Sigma_{A_i}^{(n)} \propto\sum_{\{k_j\}} C_{\{k_j\}}\prod_{j=1}^{n}\Phi_{k_j} \mkern-2.5mu\big(r^{A_i}_{j}\big) ,
\end{equation}
where $k_j$ labels a complete set of operators on the $j^{\textrm{th}}$ copy, and $r^{A_i}_{j}$ is an arbitrary point in $A_i$ on that copy. 
We have the analogous equation for $\Sigma_{A_1}^{(n)T_1}$, but the coefficients $C_{\{k_j\}}$ are replaced by $\widetilde{C}_{\{k_j\}}$. These coefficients depend on the region they pertain to, but we leave this dependence implicit.
For scalar operators we have~\cite{C13} 
\begin{equation}\label{eq:C_corr}
\begin{split}
    C_{\{k_j\}} &= \lim_{\{r_j\} \to \infty}  \left \langle \prod_{j=1}^{n} r_{j}^{\Delta_{k_j}}\Phi_{k_j}\mkern-2.0mu(r_{j})\right \rangle_{\hspace{-.15cm} \mathcal{C}_{A_1}^{(n)}},
    \end{split}
\end{equation}
and similarly for $ \widetilde{C}_{\{k_j\}}$ with a time-reversed manifold $\mathcal C_{A_1}^{(n)T_1}$. The argument generalizes to fields with spin, such as fermions and vector fields, see~\cite{agon2022tripartite}.
Crucially, since the coefficients depend on manifolds related by time-reversal symmetry, we have $\widetilde{C}_{\{k_j\}}=\pm C_{\{k_j\}}$ depending on the parity of $\{{k_j}\}$. 
We thus find 
\begin{equation}
\label{eq:Ebn_scaling_CFT}
     \mathcal{E}^b_n = \sum_{\{{k}_j\}} C_{\{{k}_j\}}( \widetilde{C}_{\{{k}_j\}}-C_{\{{k}_j\}})\ r^{-2\sum_j \Delta_{{k}_j}} + \dotsb \ .
\end{equation}
When $n\to 1$, the $C,\widetilde{C}$ coefficients reduce to groundstate one-point functions which vanish on $\mathbb{R}^{d+1}$. Therefore, $\mathcal{E}^b$ decays faster than any power, generalizing the $d=1$ results~\cite{CCT12, CCT13} to CFTs in arbitrary dimensions, including fermionic ones. In contrast, the moments $\mathcal{E}^b_n$ decay algebraically. The leading contribution corresponds to the combination of operators with the lowest sum of conformal dimensions which is odd under time reversal. Also, the non-vanishing terms in the expansion are simply $2(1-n)$ times the corresponding ones in the expansion of the R\'enyi mutual information~\cite{C13}. Finally, for regions of different shapes, $\ell^2/r^2$ in \eqref{eq:En-power} becomes $\ell_1\ell_2/r^2$, and the overall prefactor depends on the geometry ($\ell_i$ is the linear size of $A_i$). Such shape dependence 
also holds for the fermionic LN discussed below,
and its analysis represents an avenue for future research.

As an application of \eqref{eq:Ebn_scaling_CFT}, let us consider the Ising CFTs in 1d and 2d. The first primary operators are $\mathbb{I}, \sigma, \epsilon$ with respective dimensions $0<\Delta_\sigma<\Delta_\epsilon$. These fields are time-reversal symmetric; the first odd operator is the time component of the descendant $\partial_\mu \sigma$, with scaling dimension $\Delta_{\sigma}\!+\!1$. The first contribution in the expansion~\eqref{eq:Ebn_scaling_CFT} thus comes from the configuration $\{\sigma, \partial_\mu \sigma\}$ where the $\mathbb Z_2$ symmetry requires an even number of $\sigma$ fields, and the moments scale precisely as in \eqref{eq:En-power}. 

\section{Fermion entanglement}
As we did for spin Hamiltonians, we first work in the skeletal regime where each subregion contains a site hosting a single fermion mode. As we explained above, this skeletal regime faithfully yields the long-distance scaling by virtue of the OPE. Our approach will capture the non-perturbative contributions of the leading fermionic and bosonic operators in the spectrum. To obtain subleading contributions, one would increase the number of modes kept in each subregion.
The creation operators are $c^\dag_{i_1}, c^\dag_{i_2}$, and the reduced density matrix $\rho_A$ and its fermionic PT read in the occupation-number $\n_i\!=\!c^\dag_i c_i$ basis, $\{|0\rangle,c^\dag_{i_2}|0\rangle , c^\dag_{i_1}|0\rangle,c^\dag_{i_1}c^\dag_{i_2}|0\rangle\}$, 
\begin{align} \label{eq:rho-fermion}
\rho_A^{\left( T_1^f \right)} \!=\! \begin{bmatrix}
1-2\bar \n+\varrho_r^\n & 0 & 0 & \big(i\, g_r^* \big) \\
0 & \bar \n-\varrho_r^\n & g_r & 0 \\
0 & g_r^* & \bar \n-\varrho_r^\n & 0 \\
 \big(i\, g_r \big) & 0 & 0 & \varrho_r^\n
\end{bmatrix}
\end{align}
where $\bar \n=\langle \n_{i_1}\rangle$ is the average occupation, $\varrho_r^\n=\langle \n_{i_1} \n_{i_2}\rangle$ is the occupation 2-point function,  
$g_r\!=\!\langle c_{i_1}^\dag c_{i_2}\rangle$ is the fermion Green's function, and $r=|i_2-i_1|$ the separation (see SM). 
For $\rho_A$, one sets the elements  in parentheses (14 and 41) to zero, while to get $\rho_A^{T_1^f}$ we set the $g$-terms not in parentheses (23 and 32) to zero. 
The fermionic PT is seen to correspond to the regular PT $T_1^b$ accompanied by an additional multiplication by a phase~\cite{SSR17}, see also SM. 
For simplicity, we have assumed that (i) sites $i_1$ and $i_2$ are related by symmetry and (ii) the state 
conserves the fermion number, such that $\langle c_{i_1}c_{i_2}\rangle$ vanishes. 
As before, we work with Hamiltonians on the infinite lattice or with periodic boundary conditions; for open boundary conditions, the two sites should be related by reflection. 
The above two conditions hold in a great variety of cases, 
including Hamiltonians that lead to CFTs such as the Su-Schrieffer-Heeger (SSH) model at its topological phase transition~\cite{su1979solitons}, or more interestingly the interacting Hubbard model of spinless fermions on the square lattice,
$H_f=\sum_{\langle i,j\rangle} ((-1)^{s_{ij}}c_i^\dag c_j +V\n_i \n_j)$. Here, $V\geqslant0$ is a repulsion between nearest neighbors, while the phase $(\pm 1)$ in the hoppings generates a $\pi$-flux on each square plaquette so that the $V\!=\!0$ band structure has two Dirac cones.
At half-filling,~$V_c$ corresponds to a Gross-Neveu-Yukawa quantum critical point separating a Dirac semi-metal from a gapped charge density wave (CDW) phase where the occupations for the two sites in the unit cell differ~\cite{wang2014fermionic,li2015fermion}. Entanglement was previously studied for $V\leqslant V_c$ but for adjacent subregions~\cite{zhu2018entanglement}. 

\begin{figure}
    \centering
    \includegraphics[width=0.45\textwidth]{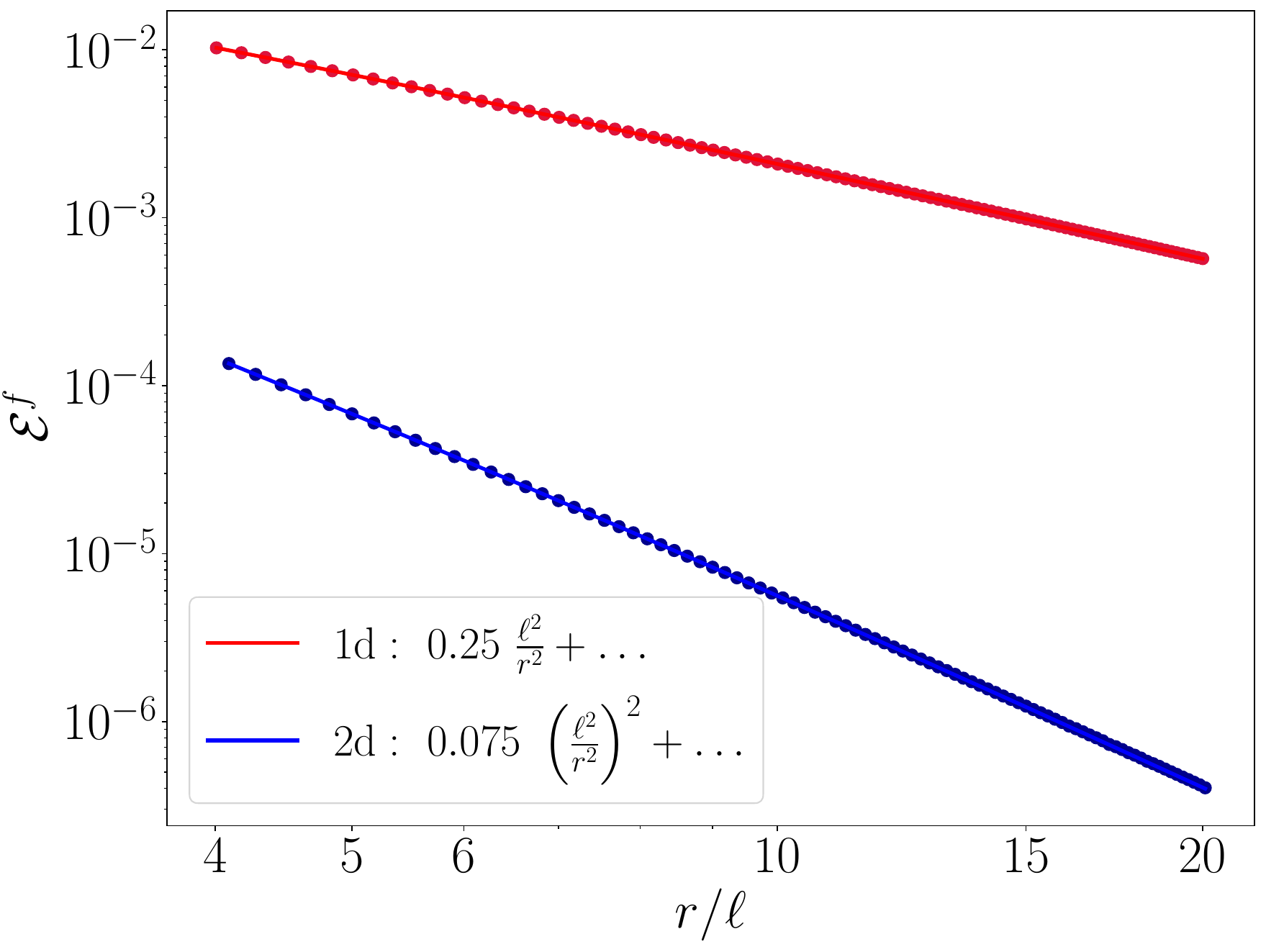}
    \caption{Fermionic LN $\mathcal E^f$ for Dirac fermions in 1d (red) and 2d (blue) versus $r/\ell$ in log-log scale. Data obtained by numerical diagonalization of the correlation matrix for $\ell=120$ (1d) and $\ell=11$ (2d). The solid lines correspond to the prediction \eqref{eq:ef} with $\Delta_f=d/2$. For $d\!=\!1$ we find $B\!=\!1/4$, in agreement with Ref.~\cite{shapourian2019twisted}, whereas for $d=2$ it is obtained from a fit.}
    \label{fig:Ef_Dirac}
\end{figure}

Using \eqref{eq:rho-fermion} we can analytically evaluate the fermionic LN. Let us work at half-filling $\bar \n\!=\! 1/2$. From the definition \eqref{eq:ln}, we get $\mathcal E^f=\log[\tfrac12 (1-4\rho_r^\n)+\tfrac12\sqrt{16 |g_r|^2+(1+4\rho_r^\n)^2} ]$ (see SM), where $\rho_r^\n=\varrho_r^\n-\bar \n^2$ is the connected correlation function. If the lattice model describes a quantum critical system, we  
have the following power-law decay at large separations: $g_r=A_f/r^{2\Delta_f}$ and $\rho_r^\n=A_\n/r^{2\Delta_\n}$. In turn, this implies the power-law decay of the fermionic LN,
\begin{align} \label{eq:ef}
    \mathcal E^f(r\gg \ell) = B\left(\frac{\ell^2}{r^2}\right)^{\!2\Delta_f},
\end{align}
where we have written the expression in a form that we conjecture 
holds true in the thermodynamic limit for a large family of CFTs, with $\ell$ being the linear size of $A_1$ and $A_2$. In contrast, the bosonic LN of \eqref{eq:rho-fermion} is readily seen to vanish for $r\geqslant r_{\rm sd}$, in agreement with our findings of the previous section. 
The leading term in \eqref{eq:ef} is always given by the fermionic scaling dimension $\Delta_f$. Interestingly, the scaling dimension $\Delta_\n$ can be smaller than $\Delta_f$ since $\n_i$ will generally have overlap with a scalar in the CFT, and the unitary bound for scalars, $(d-1)/2$, is lesser than for fermions, $d/2$. So $\mathcal E^f(r\!\gg\!\ell)$ is not determined by the operator with the lowest dimension, but rather by the lowest \emph{fermionic} operator. 
Indeed, the first terms in the expansion  are $\mathcal E^f=4|g|^2 -24|g|^4 - 16|g|^2\rho_r^\n+\dotsb$. We observe that no term coming solely from bosonic operators  appears. In contrast, for the mutual information between the two sites in the same limit, we get $I=4|g|^2+8(\rho_r^\n)^2+\dotsb$. The second term is purely bosonic, and would dominate in theories with $\Delta_\n<\Delta_f$, where it would characterize non-entangling, i.e.,\ classical, correlations. A case where this occurs is the model $H_f$ at $V_c$ where the number operators $\n_{i_1}$ and $\n_{i_2}$ are not even under the $\mathbb Z_2$ symmetry exchanging the two sites within a unit cell. As such $\rho_r^\n$ will have overlap with the CFT correlator of the order parameter field $\sigma$, which has a lower scaling dimension than the fermion~\cite{wang2014fermionic,li2015fermion,bootGN}. We also find   that the moments $\mathcal E_n^f$, defined analogously to \eqref{eq:En-power}, 
scale with the same power as $\mathcal E^f$ in \eqref{eq:ef}, see SM. 

In Fig.~\ref{fig:Ef_Dirac}, we test \eqref{eq:ef} for free Dirac fermion CFTs in $d=1,2$. The Gaussian groundstates
allow us to reach large system sizes $\ell^d\gg1$ on the lattice. The scaling agrees perfectly with \eqref{eq:ef} for the dimensions $\Delta_f=d/2$.

\section{Outlook}
We showed that at large separations the bosonic LN $\mathcal E^b$ decays faster than any power, whereas the fermionic LN $\mathcal E^f$ decays algebraically for fermionic quantum critical systems. There is thus no distillable long-range entanglement in spin (bosonic) systems, in stark contrast with fermionic ones whose entanglement is infinitely more robust.
It would be interesting to understand the scaling function for general CFTs, 
but the answer escapes the present expansion of twist operators. Further, it would be of interest to adapt the twist expansion to show the algebraic decay of the fermionic LN \eqref{eq:ef} in the continuum. For boson/spin systems, a fundamental question arises: since quantum critical groundstates do not possess such long-range entanglement, what physically relevant quantum states do? A natural setting would be to study non-equilibrium states.
A result in this direction was obtained for quantum circuits at the 1d measurement-driven phase transition, where algebraic decay was observed~\cite{PRXQuantum,shi2021entanglement}. Our study motivates charting the space of long-range entangled states. \vspace{0.5cm}

\section*{Acknowledgements}
We thank V.~Alba, P.~Calabrese, H.~Casini, and E.~Tonni for useful discussions. G.P. holds an FRQNT Postdoctoral Fellowship, and acknowledges support from the Mathematical Physics Laboratory of the Centre de Recherches Math\'ematiques (CRM). W.W.-K.\ is supported by a grant from the Fondation Courtois, a Chair of the Institut Courtois, a Discovery Grant from NSERC, and a Canada Research
Chair.

\providecommand{\href}[2]{#2}\begingroup\raggedright\endgroup

\appendix

\onecolumngrid
\clearpage
\begin{center}
\textbf{\large Supplemental Material: Entanglement negativity between separated regions in quantum critical systems} \\[.4cm]
Gilles Parez$^{1, 2}$ and William Witczak-Krempa$^{1,2,3}$ \\[.2cm]
{\it $^{1}$D\'epartement de Physique, Universit\'e de Montr\'eal, Montr\'eal, QC H3C
3J7, Canada}\\
{\it $^{2}$Centre de Recherches Math\'ematiques, Universit\'e de Montr\'eal, Montr\'eal, QC H3C 3J7, Canada}\\
{\it $^{3}$Institut Courtois, Universit\'e de Montr\'eal, Montr\'eal, QC H2V 0B3, Canada}

\end{center}

\setcounter{equation}{0}
\setcounter{figure}{0}
\setcounter{table}{0}
\makeatletter
\renewcommand{\theequation}{S\arabic{equation}}
\renewcommand{\thefigure}{S\arabic{figure}}
\renewcommand{\bibnumfmt}[1]{[S#1]}

\medskip
\section*{A. TWO-SPIN DENSITY MATRIX} \label{sec:app}

We first show how to obtain the matrix elements of the two-spin reduced density matrix \eqref{eq:rho}. In full generality, the density matrix reads
\begin{equation}
    \rho_A = \sum_{a_{i_1},a_{i_2}=x,y,z,0}\rho_{a_{i_1}a_{i_2}} \ S_{i_1}^{a_{i_1}}  S_{i_2}^{a_{i_2}}
\end{equation}
where $S^{x,y,z}$ are the spin-1/2 operators and $S^0=\frac 12\mathbb{I}_2$. The coefficients $\rho_{a_{i_1}a_{i_2}}$ are obtained as 
\begin{equation}\label{eq:rho_coeff}
    \rho_{a_{i_1}a_{i_2}} =4 \Tr(\rho_A S_{i_1}^{a_{i_1}}  S_{i_2}^{a_{i_2}})\ .
\end{equation}
 Let us now compute the matrix elements of $\rho_A$ in the spin basis given in the main text. We have 
 \begin{equation}
     \begin{split}
         [\rho_A]_{11} &= \langle \downarrow \downarrow \! |\rho_A| \!  \downarrow \downarrow \rangle \\
         &=\frac14 ( \rho_{00}+\rho_{zz}-\rho_{0z}-\rho_{z0}).
     \end{split}
 \end{equation}
 With Eq.~\eqref{eq:rho_coeff} we find $\rho_{00}=1$, $\rho_{zz} =4 \Tr(\rho_A S_{i_1}^zS_{i_2}^z) =4\langle S_{i_1}^zS_{i_2}^z\rangle \equiv 4 \varrho_r^z $, and similarly $\rho_{0z}=\rho_{z0} = 2 m_z$, giving $ [\rho_A]_{11} = 1/4-m_z+\varrho_r^z$, as expected. The computation of all other matrix elements follows the exact same reasoning.

\subsection*{Bosonic negativity}
The eigenvalues of the reduced density matrix of two spins \eqref{eq:rho} are
\begin{equation} \label{eq:lam}
    \begin{split}
        \lambda_1 &= \tfrac14 - x - y -z \\ 
    \lambda_2 &= \tfrac14 + x + y - z  \\
    \lambda_3 &= \tfrac14 + z - \sqrt{m_z^2 +(x-y)^2} \\ 
    \lambda_4 &= \tfrac14 + z + \sqrt{m_z^2 +(x-y)^2},
    \end{split}
\end{equation}
where we introduced the following notation for the spin correlators
\begin{align}
    x=\varrho_r^x\,, \quad y=\varrho_r^y\,, \quad  z=\varrho_r^z  .
\end{align}
As explained in the main text, the bosonic PT transforms $y\to-y$ in the original density matrix. The eigenvalues of $\rho_A^{T_1^b}$, which is Hermitian, are then
\begin{align} \label{eq:lam-pt}
    \tilde\lambda_i =\lambda_i\big|_{y\to -y}.
\end{align}
The bosonic LN can then be computed via $\mathcal E^b =\log \sum_i|\tilde\lambda_i|$. 

Let us now discuss the case of the $d$-dimensional quantum Ising model $H=-\sum_{\langle i,j\rangle} S_i^x S_j^x -h\sum_i S_i^z$, with arbitrary $h\geqslant 0$. (The case $h<0$ can be obtained by symmetry, and the same conclusions hold.) 
The correlation function $y=\langle S_{i_1}^y S_{i_2}^y\rangle$ decays to zero at large $r=|i_2-i_1|$; at the quantum critical point $h=h_c$, this decay is algebraic. Hence, for sufficiently large $r$, the eigenvalues \eqref{eq:lam-pt} become
\begin{align}
    \tilde\lambda_i =\lambda_i^\infty + \delta\mkern-1mu\lambda_i 
\end{align}
with the deviation being parametrically small in the separation, $|\delta\mkern-1mu\lambda_i/\lambda_i^\infty|\ll 1$. Here, $\lambda_i^\infty$ denotes the $r=\infty$ value of the $\rho_A$ eigenvalues $\lambda_i$ in Eq.~\eqref{eq:lam}. 
It can be seen that the eigenvalues $\lambda_i$ at $r=\infty$ are strictly positive for $0<h<\infty$. For instance, $\lambda_1^\infty=\tfrac14-m_x^2-m_z^2$, where $m_x=\langle S_i^x\rangle$ is the 1-point function of the order parameter.
The Heisenberg uncertainty principle constrains the 1-point functions to obey $m_x^2+m_z^2\leqslant 1/4$. 
Thus, this eigenvalue 
only vanishes at the two extreme points in the phase diagram: $h=0$ where $m_x=\pm 1/2$, and $h=\infty$ where $m_z=1/2$. 
For instance, at the critical point the order parameter vanishes $m_x=0$, while $m_z<1/2$ since the state is not fully polarised along the transverse field direction. 
We conclude that there exists a finite critical separation $r_{\rm sd}$ at which  $\tilde\lambda_i$ become strictly positive, and so
\begin{align}
    \mathcal E^b(r\geqslant r_{\rm sd})=0,
\end{align}
implying that the 2-spin state $\rho_A$ becomes separable beyond a critical separation. We note this separability also holds at the extremes $h=0,\infty$ since the state $\rho_A$ is then manifestly separable.
Indeed, one can check that while some eigenvalues $\tilde\lambda_i$ vanish, none become negative.

\subsection*{Bosonic negativity moments} 
Although the density matrix becomes separable it is not invariant under PT, and we can quantify this non-invariance via the moments
\begin{align}
    \mathcal E_n^b &= \log \frac{\Tr\big(\rho_A^{T_1^b}\big)^n}{\Tr \rho_A^n} =\log \frac{\sum_i \tilde\lambda_i^n}{\sum_i \lambda_i^n}.
\end{align}
Using the above results for the eigenvalues, we find for large separations at criticality 
\begin{equation}
    \begin{split}
        \mathcal E_n^b &= b_n\, xy + \dotsb \\
    &= b_n \frac{A_x A_y}{r^{2(2\Delta_\sigma+1)}}+ \dotsb
    \end{split}
\end{equation}
in agreement with the general CFT scaling obtained in the main text.
For $n=3,4,5$, the $b_n$ coefficients read
\begin{equation}
    \begin{split}
         b_3 &= \frac{768\,m_z^2}{(1+12m_z^2)^2} \\
    b_4 &=  \frac{4096\,m_z^2}{(1+24m_z^2+16m_z^4)^2} \\
    b_5 &= \frac{2560 \, m_z^2 (5+ 8 m_z^2+ 16 m_z^4)}{(1+ 40
   m_z^2+80 m_z^4)^2}.
    \end{split}
\end{equation}

\section*{B. TWO-FERMION DENSITY MATRIX} \label{sec:app2}

We compute the matrix element $[\rho_A]_{ij}$ of the two-fermions density matrix \eqref{eq:rho-fermion} in the basis given in the main text. We begin with the off-diagonal elements,
\begin{equation}
\begin{split}
    [\rho_A]_{23} &= \langle 0 | c_{i_2}\rho_A c_{i_1}^\dagger |0\rangle \\
    &= \langle 0 | c_{i_2}\rho_A c_{i_1}^\dagger |0\rangle +\langle 0 | c_{i_2} c_{i_2}\rho_A c_{i_1}^\dagger  c_{i_2}^\dagger|0\rangle +\langle 0 | c_{i_1} c_{i_2}\rho_A c_{i_1}^\dagger c_{i_1}^\dagger |0\rangle +\langle 0 | c_{i_2}c_{i_1}c_{i_2}\rho_A c_{i_1}^\dagger c_{i_1}^\dagger c_{i_2}^\dagger|0\rangle \\
    &= \Tr(\rho_A c_{i_1}^\dagger c_{i_2}) \\
    &= \langle  c_{i_1}^\dagger c_{i_2}\rangle \equiv g_r ,
    \end{split}
\end{equation}
and similarly we have $ [\rho_A]_{32}=g_r^*$. On the second line, each of the three last terms vanishes because fermionic operators satisfy the exclusion principle $c^2 = (c^\dagger)^2 = 0$. Moreover, we used the cyclic property of the trace from the second to the third line. We compute the diagonal elements using similar arguments. We start with $ [\rho_A]_{44}$,
\begin{equation}
\label{eq:44}
    \begin{split}
         [\rho_A]_{44} &= \langle 0 | c_{i_2}c_{i_1} \rho_A c_{i_1}^\dagger c_{i_2}^\dagger  | 0\rangle \\
         &= \Tr( \rho_A c_{i_1}^\dagger c_{i_1} c_{i_2}^\dagger c_{i_2} ) \\
         &= \langle \n_{i_1}\n_{i_2}\rangle \equiv \varrho_r^\n,
    \end{split}
\end{equation}
where we used the standard anticommutation relation $\{c_{i_1},c_{i_2}^{(\dagger)}\}=0$ to obtain the second line. Next,
\begin{equation}
    \begin{split}
         [\rho_A]_{33} &=  \langle 0 |c_{i_1} \rho_A c_{i_1}^\dagger  | 0\rangle \\
         &= \Tr(\rho_A c_{i_1}^\dagger c_{i_1}) -\langle 0 | c_{i_2}c_{i_1} \rho_A c_{i_1}^\dagger c_{i_2}^\dagger  | 0\rangle\\
         &= \langle \n_{i_1}\rangle - \langle \n_{i_1}\n_{i_2}\rangle \equiv \bar{\n} - \varrho_r^\n,
    \end{split}
\end{equation}
where we used Eq.~\eqref{eq:44} from the second to the third line. The computation of $[\rho_A]_{22}$ is identical. Finally, using $\Tr(\rho_A)=1$ we find  $[\rho_A]_{11} = 1-2\bar{\n} + \varrho_r^\n$. All the other matrix elements vanish. 

\subsection*{Fermionic partial transpose}

In this section we give the action of fermionic PT $T_1^f$ in the number-occupation basis. We consider basis states $|\alpha_1\rangle|\beta_2\rangle$ for the system $A_1A_2$ of the form 
\begin{equation}
    |\alpha_1\rangle = \prod_{j\in A_1} (c_j^\dagger)^{n_j} |0\rangle, \quad  |\beta_2\rangle = \prod_{j\in A_2} (c_j^\dagger)^{n_j} |0\rangle, 
\end{equation}
where $n_j=0,1$. The occupation numbers of $|\alpha_1\rangle$ and $|\beta_2\rangle$ are $f_1 = \sum_{j\in A_1}n_j$ and $f_2 = \sum_{j\in A_2}n_j$, respectively. The fermionic PT reads \cite{SSR17}
\begin{equation}
    T_1^f\left( |\alpha_1\rangle|\beta_2\rangle\, \langle\widetilde\alpha_1|\langle\widetilde\beta_2|\right)=(-1)^{\phi(\{n_j\},\{\tilde n_j\})}|\widetilde\alpha_1\rangle|\beta_2\rangle\, \langle\alpha_1|\langle\widetilde\beta_2|
\end{equation}
where the phase is 
\begin{equation}
   \phi(\{n_j\},\{\tilde n_j\}) = \frac{f_1(f_1+2)}{2}+ \frac{\tilde f_1(\tilde f_1+2)}{2} + f_2 \tilde f_2 + f_1 f_2 + \tilde f_1 \tilde f_2 + (f_1+f_2)(\tilde f_1 + \tilde f_2). 
\end{equation}

As an example, in the case of two fermionic modes we have 
\begin{equation}
     T_1^f( c_{i_2}^\dagger |0\rangle \langle 0| c_{i_1}  ) = i \ c_{i_1}^\dagger c_{i_2}^\dagger|0\rangle \langle  0|,
\end{equation}
which implies $[\rho_A^{T_1^f}]_{41} = i [\rho_A]_{23} $ as in \eqref{eq:rho-fermion}. 

\subsection*{Fermionic negativity}
Working at half-filling $\bar{\rm n}=1/2$, the eigenvalues of the reduced density matrix of two fermions \eqref{eq:rho-fermion} are
\begin{equation}\label{eq:lam-f}
    \begin{split}
         \lambda_1=\lambda_2 &= \tfrac14 + \rho^{\rm n} \\ 
    \lambda_3 &= \tfrac14 -  \rho^{\rm n}  - |g| \\ 
    \lambda_4 &= \tfrac14 -  \rho^{\rm n}  + |g|
    \end{split}
\end{equation}
where $\rho^{\rm n}=\varrho^{\rm n}-\bar \n^2$ is the connected correlation function; we have left the $r$-dependence implicit. The eigenvalues $\hat\lambda_i$ of $\rho_A^{T_1^f}(\rho_A^{T_1^f})^\dagger$ are
\begin{equation}
    \begin{split}
         \hat\lambda_1= \hat\lambda_2 &= \frac{1}{16}(1 - 4\rho^{\rm n})^2, \\ 
          \hat\lambda_3= \hat\lambda_4 & =  |g|^2+\frac{1}{16}(1+4\rho^\n)^2.
    \end{split}
\end{equation}
One can then readily obtain the fermionic LN as
\begin{equation}
    \begin{split}
        \mathcal E^f &= \log\sum_i (\hat\lambda_i)^{1/2} \\
    &= \log\left[\frac12 (1-4\rho^\n)+ \frac12\sqrt{16 |g|^2+(1+4\rho^\n)^2} \right]
    \end{split}
\end{equation}
where we 
used $|\langle \n_{i_1}\n_{i_2}\rangle|\leqslant 1/2$, which follows from the Cauchy-Schwartz inequality.

\subsection*{Fermionic negativity moments}
As for the bosonic LN, we can introduce normalised moments of the fermionic PT
\begin{align}
    \mathcal E_n^f = \log \frac{\Tr \left(\rho_A^{T_1^f}\big(\rho_A^{T_1^f}\big)^{\!\dag}\right)^{\mkern-4mu n} }{\Tr \rho_A^{2n}}  
    =\log \frac{\sum_i (\hat\lambda_i)^{n} }{\sum_i \lambda_i^{2n} }. 
\end{align}
The analytic continuation of the integer values to $n=1/2$ gives the LN $\mathcal E^f$. These are the quantities that one needs to evaluate in the replica method~\cite{SSR17}. The first non-trivial moment, $n\!=\!2$ ($n\!=\!1$ identically 
gives zero), reads at large separations
\begin{align}
   \mathcal E_2^f= -32 |g|^2 + 1024 |g|^4 + 512 |g|^2 \rho^{\rm n} + \dotsb 
\end{align}
where we see that all terms contain the fermionic Green's function $g$.
The same expansion form holds for other values of $n$ but with different coefficients. For example, the leading term is in general
\begin{align}
    \mathcal E_n^f= 16n(1-n)|g|^2+\dotsb 
\end{align}
which indeed yields $\mathcal E^f$ as $n\to1/2$, and vanishes when $n=1$.
Using the large-$r$ scaling $g=A_f/r^{2\Delta_f}$ and $\rho^\n=A_\n/r^{2\Delta_\n}$, we thus find the same scaling as for $\mathcal E^f$:
\begin{align} 
    \mathcal E_n^f(r\gg \ell) = f_{n} \left(\frac{\ell^2}{r^2}\right)^{\!2\Delta_f} + \dotsb
\end{align}
where we have written the expression in a form that we conjecture
holds true in the thermodynamic limit for a large family of CFTs, with $\ell$ being the linear size of $A_1$ and $A_2$.

\subsection*{Mutual information}
In order to evaluate the mutual information between the 2 sites,
\begin{align}
    I(A_1,A_2)= S(A_1)+S(A_2)-S(A_1 A_2),
\end{align}
we first need the reduced density matrix of a single site, say $A_1=\{i_1\}$:
\begin{align}
    \rho_{A_1}=\Tr_{A_2}\rho_{A_1A_2}=(1-\langle {\rm n}_{i_1}\rangle)|0\rangle\langle 0| + \langle {\rm n}_{i_1}\rangle |1\rangle\langle 1|
\end{align}
which only depends on the average occupation of site $i_1$. Working at half-filling, $\langle {\rm n}_{i_1}\rangle=\bar \n=1/2$, we get the entanglement entropy 
\begin{align}
    S(A_1)=-\Tr(\rho_{A_1}\log \rho_{A_1})=\log 2,
\end{align}
and $S(A_2)$ takes the same value by symmetry. The entanglement entropy of $A\!=\! A_1A_2$ can be readily obtained from the eigenvalues \eqref{eq:lam-f}:
\begin{align}
    S(A_1 A_2)=-\sum_i \lambda_i \log\lambda_i.
\end{align}
Finally, expanding $I(A_1,A_2)$ at large separations, we get the result quoted in the main text. 

\end{document}